\documentclass[12pt,twoside]{article}
\usepackage[english]{babel}
\usepackage{graphicx,rotating,epsfig}
\usepackage{a4p}
\usepackage{xspace}

\def\ra{\rightarrow} 
\def\GeV  {\ensuremath{\mathrm{ Ge\kern -0.1em V } }}
\def\GeVc2{\ensuremath{\mathrm{ Ge\kern -0.1em V }\kern -0.2em /c^2 }}
\def\MeVc2{\ensuremath{\mathrm{ Me\kern -0.1em V }\kern -0.2em /c^2 }}
\newcommand{\MT}{\ensuremath{{\mathrm{m}}_{\mathrm{top}}}}
\newcommand{\MTll}{\ensuremath{\MT^{\mathrm{di\mbox{-}l}}}}
\newcommand{\MTlj}{\ensuremath{\MT^{\mathrm{l\mbox{+}j}}}}
\newcommand{\MTjj}{\ensuremath{\MT^{\mathrm{all\mbox{-}j}}}}

\newcommand{\pb}{\ensuremath{\mathrm{pb}^{-1}}}
\newcommand{\fb}{\ensuremath{\mathrm{fb}^{-1}}}
\newcommand{\ttbar}{\ensuremath{t\overline{t}}}
\newcommand{\WbWb}{\ensuremath{W^+ b W^- \overline{b}}}
\newcommand{\ljt}{\ensuremath{\ell\nu b q q^{\prime} \overline{b}}}
\newcommand{\had}{\ensuremath{q q^{\prime} b q q^{\prime} \overline{b}}}
\newcommand{\dil}{\ensuremath{\ell^{+}\nu b\ell^{-}\overline{\nu}\overline{b}}}
\newcommand{\ttljt}{\ensuremath{\ttbar\ra\WbWb\ra\ljt}}
\newcommand{\ttdil}{\ensuremath{\ttbar\ra\WbWb\ra\dil}}
\newcommand{\tthad}{\ensuremath{\ttbar\ra\WbWb\ra\had}}
\newcommand{\RunI}{\hbox{Run\,I}}
\newcommand{\RunII}{\hbox{Run\,II}}
\newcommand{\measStatSyst}[3]{\ensuremath{#1 \pm #2~(\textrm{stat}) \pm #3~(\textrm{syst})}\xspace}
\newcommand{\gevcc}[1]  {\ensuremath{#1~\mathrm{GeV}/c^{2}}}

\begin{document}

\begin{center}
  {\LARGE FERMI NATIONAL ACCELERATOR LABORATORY}
\end{center}

\begin{flushright}
       FERMILAB-TM-2466-E \\  
       TEVEWWG/top 2010/07 \\
       CDF Note 10210 \\
       D\O\ Note 6090 \\
       \vspace*{0.05in}
       July 2010 \\
\end{flushright}

\vskip 1cm

\begin{center}
  {\LARGE\bf 
    Combination of CDF and D\O\ results 
    on the mass of the top quark using up to \boldmath{$5.6\:\fb$} of data\\
  }
  \vfill
  {\Large
    The Tevatron Electroweak Working Group\footnote{The Tevatron Electroweak 
    Working Group can be contacted at tev-ewwg@fnal.gov.\\  
    \hspace*{0.20in} More information can
    be found at {\tt http://tevewwg.fnal.gov}.} \\
    for the CDF and D\O\ Collaborations\\
  }
\end{center}
\vfill
\begin{abstract}
\noindent
  We summarize the top-quark mass measurements from the CDF and
  D\O\ experiments at Fermilab.  We combine published
  \RunI\ (1992--1996) measurements with the most precise published and preliminary
  \RunII\ (2001-present) measurements using up to $5.6~\fb$ of data.
 Taking uncertainty correlations properly into account, and
 adding in quadrature the statistical and systematic uncertainties, the resulting
  preliminary Tevatron average mass of the top quark is $\MT = \gevcc{173.3 \pm 1.1}$.
  
 
\end{abstract}

\vfill



\section{Introduction}
\label{sec:intro}

This note reports the Tevatron average top-quark mass obtained by
combining the most precise published and preliminary measurements of the top-quark mass, \MT. 

The experiments CDF and D\O, taking data at the Tevatron
proton-antiproton collider located at the Fermi National Accelerator
Laboratory, have made several direct experimental measurements of the
top-quark mass, \MT.  The pioneering measurements were based on about
$100~\pb$ of \RunI\ data~[1 - 12]
  collected from 1992 to 1996,
and include results from the \tthad\ (all-j), \ttljt\ (l+j)\footnote{Here $\ell=e$ or $\mu$.  Decay 
channels with explicit tau lepton identification are presently under 
study and are not yet used for measurements of the top-quark mass. Decays with $\tau \to e, \mu$ are included in the direct $W \to e$ and $W \to \mu$ channels.}, and 
\ttdil\ (di-l) decay channels.   
Several more measurements have been performed in \RunII\ (2001 - present) in all decay modes.
The \RunII\ measurements considered here are the most recent results in the 
l+j, di-l,  and all-j channels using $1.9-5.6~\fb$ of data and improved 
analysis techniques~[13 -- 20].

With respect to the March 2009 combination\,\cite{Mtop-tevewwgWin09}, the preliminary \RunII\ CDF measurement in the l+j channel has been updated using 5.6\,fb$^{-1}$ of data, and improved analysis technique\,\cite{Mtop2-CDF-l+j-new}. The now published \RunII\ CDF measurement in the di-l channel\,\cite{Aaltonen:2008bd} has been substituted with a more precise preliminary result that uses 4.8\,fb$^{-1}$ of data\,\cite{Mtop2-CDF-di-l-new}. Also, the \RunII\ measurements in the all-j channel, and the measurement based on charged particle tracking have been published\,\cite{Mtop2-CDF-all-j-new, Mtop2-CDF-trk-new}. The D\O\ \RunII\ measurements are unchanged.
The Tevatron average top-quark mass is thus obtained by combining five published
\RunI\ measurements~\cite{Mtop1-CDF-di-l-PRLb, Mtop1-CDF-di-l-PRLb-E,
  Mtop1-D0-di-l-PRD, Mtop1-CDF-l+j-PRD, Mtop1-D0-l+j-new1,
  Mtop1-CDF-all-j-PRL} with two published \RunII\ CDF results~\cite{Mtop2-CDF-all-j-new,
Mtop2-CDF-trk-new}, two preliminary \RunII\ CDF
results~\cite{Mtop2-CDF-di-l-new, Mtop2-CDF-l+j-new} and two preliminary
\RunII\ D\O\ results~[18 -- 20].
The combination takes into account the
statistical and systematic uncertainties and their correlations using
the method of Refs.~\cite{Lyons:1988, Valassi:2003} and
supersedes previous
combinations~\cite{Mtop1-tevewwg04,Mtop-tevewwgSum05,
  Mtop-tevewwgWin06,Mtop-tevewwgSum06, Mtop-tevewwgWin07, Mtop-tevewwgWin08, 
  Mtop-tevewwgSum08, Mtop-tevewwgWin09}. 
The definition and evaluation of the systematic uncertainties and the understanding of the
correlations among channels, experiments, and Tevatron runs, is the outcome of many years of 
joint work between the CDF and D\O\ collaborations.



The input measurements and uncertainty categories used in the combination are 
detailed in Sections~\ref{sec:inputs} and~\ref{sec:uncertainty}, respectively. 
The correlations used in the combination are discussed in 
Section~\ref{sec:corltns} and the resulting Tevatron average top-quark mass 
is given in Section~\ref{sec:results}.  A summary and outlook are presented
in Section~\ref{sec:summary}.
 
\section{Input Measurements}
\label{sec:inputs}

For this combination eleven measurements of \MT\ are used: five
published \RunI\ results, two published \RunII\ results, and four preliminary
\RunII\ results, all reported in Table~\ref{tab:inputs}.  In general,
the \RunI\ measurements all have relatively large statistical
uncertainties and their systematic uncertainties are dominated by the
total jet energy scale (JES) uncertainty.  In \RunII\, both CDF and
D\O\ take advantage of the larger \ttbar\ samples available and employ
new analysis techniques to reduce both these uncertainties.  In
particular, the \RunII\ D\O\ analysis in the l+j channel and the 
\RunII\ CDF analyses in the l+j and all-j channels 
constrain the response of light-quark jets using the kinematic information from $W\ra
qq^{\prime}$ decays ({\em in situ} calibration). Residual JES uncertainties associated with
$p_{T}$ and $\eta$ dependencies as well as uncertainties specific to
the response of $b$-jets are treated separately. The
\RunII\ CDF and D\O\ di-l measurements and the CDF measurement of   
Ref. \cite{Mtop2-CDF-trk-new} use a JES determined from external
calibration samples.  Some parts of the associated uncertainty are
correlated with the \RunI\ JES uncertainty as noted below.
\vspace*{0.10in}

\begin{table}[t]
\caption[Input measurements]{Summary of the measurements used to determine the
  Tevatron average $\MT$.  Integrated luminosity ($\int \mathcal{L}\;dt$) has units in
  \fb, and all other numbers are in $\GeVc2$.  The uncertainty categories and 
  their correlations are described in the Sec.\,\ref{sec:uncertainty}.  The total systematic uncertainty 
  and the total uncertainty are obtained by adding the relevant contributions 
  in quadrature. The new measurements utilized here are the two CDF preliminary results in the last column.}
\label{tab:inputs}
\begin{center}
\renewcommand{\arraystretch}{1.30}
{\small
\begin{tabular}{l|ccccc|cc|cccc} 
\hline \hline
       & \multicolumn{5}{c|}{{\RunI} published} 
       & \multicolumn{2}{c|}{{\RunII} published} 
       & \multicolumn{4}{c}{{\RunII} preliminary}  \\ 
       & \multicolumn{3}{c}{ CDF } 
       & \multicolumn{2}{c}{ D\O\ }
       & \multicolumn{2}{|c|}{ CDF }
       & \multicolumn{2}{c}{ D\O\ }
       & \multicolumn{2}{c}{ CDF }
        \\
                                      &     all-j &      l+j   & di-l  &           l+j &    di-l  &       all-j  &     trk   &       l+j        &     di-l    &       l+j     &       di-l   \\
\hline
$\int \mathcal{L}\;dt$ &      0.1 &     0.1  &     0.1  &     0.1  &     0.1 &       2.9 &          1.9 &      3.6     &        3.6 &          5.6 &       4.8 \\
\hline                         
Result                          & 186.0 &  176.1 & 167.4 & 180.1 & 168.4 &   174.80 &  175.30 & 173.75 &  174.66 &  173.00 & 170.56  \\
\hline                         
iJES                             &        -    &          - &         - &           - &          - &     1.64 &              - &        0.47 &              - &       0.58 &             -  \\
aJES                            &         -    &         - &           - &         - &             - &       -  &              - &        0.91 &       1.32 &             - &              - \\
bJES                            &   0.6    &      0.6 &     0.8 &       0.7 &      0.7 &     0.21 &         0.0 &        0.07 &       0.26 &      0.26 &        0.35 \\
cJES                            &   3.0    &      2.7 &      2.6 &      2.0 &      2.0 &      0.49 &      0.60 &          0.0 &          0.0 &      0.27 &        2.01 \\
dJES                            &   0.3    &      0.7 &      0.6 &      0.0 &      0.0 &      0.08 &        0.0 &        0.84 &       1.46 &      0.01 &        0.64 \\
rJES                             &   4.0    &      3.4 &      2.7 &      2.5 &      1.1 &      0.21 &      0.10 &          0.0 &         0.0 &      0.41 &        1.98 \\
LepPt                            &   0.0    &       0.0 &     0.0 &      0.0 &      0.0 &             - &      1.10 &        0.18 &       0.32 &       0.14 &       0.31  \\
Signal                          &   1.8    &       2.6 &     2.8 &      1.1 &      1.8 &      0.23 &      1.60 &        0.45 &      0.65 &       0.21 &        0.36 \\
Backgd                        &   1.7    &       1.3 &     0.3 &      1.0 &      1.1 &      0.35 &      1.60 &        0.08 &       0.08 &      0.34 &        0.27 \\
Fit                                 &   0.6    &       0.0 &     0.7 &       0.6 &     1.1 &      0.67 &      1.40 &         0.21 &      0.51 &      0.10 &        0.05 \\
MC                               &   0.8    &       0.1 &     0.6 &      0.0 &      0.0 &      0.31 &      0.60 &        0.58 &       1.00 &      0.37 &        0.57 \\
UN/MI                          &       -    &            - &          - &      1.3 &      1.3 &           - &             - &            -    &             -  &              - &               -  \\
CR                               &      -      &        -     &          -&           - &          - &     0.41 &      0.40 &         0.41 &       0.41 &      0.37 &       0.61  \\
MHI                             &      -      &           - &           - &          - &          - &     0.17 &      0.70 &         0.05 &        0.00 &      0.10 &       0.27  \\
\hline                         
Syst                             &   5.7    &        5.3 &     4.9 &       3.9 &      3.6 &     1.99 &      3.10 &         1.60 &     2.43 &      1.06 &        3.09 \\
Stat                              &  10.0  &         5.1 &  10.3 &       3.6 &    12.3 &    1.70 &      6.20 &          0.83 &     2.92 &      0.65 &        2.19 \\
\hline                         
Total                             &  11.5 &        7.3 &  11.4 &        5.3 &     12.8 &   2.61 &      6.94 &           1.80 &     3.80 &      1.24 &        3.79 \\ 
\hline
\hline
\end{tabular}
}
\end{center}
\label{tab:inputs}
\end{table}


The D\O\ Run~II l+j analysis uses the JES determined from the
external calibration derived from $\gamma$+jets events as an
additional Gaussian constraint to the {\em in situ} calibration. Therefore
the total resulting JES uncertainty is split into one part emerging from the 
{\em in situ} calibration and another part emerging from the external calibration.

To do that, the measurement without external
JES constraint has been combined iteratively with a pseudo-measurement
using the method of Refs.~\cite{Lyons:1988, Valassi:2003} 
which uses only the external calibration in a way that the combination give the actual total JES uncertainty. 
The splitting obtained in this way is used to assess the statistical part of the JES uncertainty, and the part of 
the JES uncertainty coming from the external calibration constraint~\cite{Mtop2-D0-comb}.

The analysis technique developed by CDF and referred to as ``trk" 
uses both the mean
decay-length from $b$-tagged jets and the mean lepton transverse momentum
to determine the top-quark mass in l+j candidate events.
While the statistical sensitivity is not as good as the more
traditional methods, this technique has the advantage that since it
uses primarily tracking information, it is almost entirely independent of
JES uncertainties.  As the statistics of this sample continue to
grow, this method is expected to offer a cross-check of the top-quark mass
largely independent of the dominant JES systematic uncertainty.  
The statistical correlation between an earlier version of the trk analysis and a
traditional \RunII\ CDF l+j measurement was
studied using Monte Carlo signal-plus-background pseudo-experiments
which correctly account for the sample overlap and was found to be
consistent with zero (to within $<$1\%) independent of the assumed
top-quark mass.
\vspace*{0.10in}

The D\O\ \RunII\ l+j result is a combination of the 
published Run~IIa (2002--2005) measurement ~\cite{Mtop2-D0-l+ja-final} with 1 fb$^{-1}$ 
of data and the preliminary result obtained with 2.6 fb$^{-1}$ Run~IIb (2006--2007) dataset.   

The D\O\ \RunII\ di-l result is itself a combination of two results
using different techniques but partially overlapping dilepton 
datasets~\cite{Mtop2-D0-di-l-mar09-1,Mtop2-D0-di-l-jul08-2}.
\vspace*{0.10in}

Table~\ref{tab:inputs} also lists the uncertainties of the results,
subdivided into the categories described in the next Section.  The
correlations between the inputs are described in
Section~\ref{sec:corltns}.


\section{Uncertainty Categories}
\label{sec:uncertainty}

We employ the same uncertainty categories as used for the previous Tevatron
average~\cite{Mtop-tevewwgWin09}.  
They are divided such that sources of systematic uncertainty that share the same or similar origin are combined.
For example, the ``Signal'' category
discussed below includes the uncertainties from initial state ratiation (ISR), 
final state radiation (FSR), and parton density functions (PDF)---all of which affect the modeling of the \ttbar\ signal.  Some
systematic uncertainties have been broken down into multiple
categories in order to accommodate specific types of correlations.
For example, the jet energy scale (JES) uncertainty is subdivided
into six components in order to more accurately accommodate our
best estimate of the relevant correlations.  
\vspace*{0.10in}

\begin{description}
  \item[Statistics:] The statistical uncertainty associated with the
    \MT\ determination.
 \item[iJES:] That part of the JES uncertainty which originates from
   {\em in situ} calibration procedures and is uncorrelated among the
   measurements.  In the combination reported here, it corresponds to
   the statistical uncertainty associated with the JES determination
   using the $W\ra qq^{\prime}$ invariant mass in the CDF \RunII\
   l+j and all-j measurements and D\O\ Run~II l+j
   measurement. Residual JES uncertainties arising from effects
   not considered in the {\em in situ} calibration are included in other
   categories.
  \item[aJES:] That part of the JES uncertainty which originates from
    differences in detector electromagnetic over hadronic ($e/h$) response 
    between $b$-jets and light-quark
    jets. This category also includes uncertainties associated with the 
    jet identification and resolution, trigger and $b$-jets tagging.    
    It is specific to the D\O\ \RunII\ measurements and is
    uncorrelated with the D\O\ \RunI\ and CDF measurements. 
  \item[bJES:] That part of the JES uncertainty which originates from
    uncertainties specific to the modeling of $b$-jets and which is correlated
    across all measurements.  For both CDF and D\O\ this includes uncertainties 
    arising from 
    variations in the semileptonic branching fractions, $b$-fragmentation 
    modeling, and differences in the color flow between $b$-jets and light-quark
    jets.  These were determined from \RunII\ studies but back-propagated
    to the \RunI\ measurements, whose rJES uncertainties (see below) were 
    then corrected in order to keep the total JES uncertainty constant.
  \item[cJES:] That part of the JES uncertainty which originates from
    modeling uncertainties correlated across all measurements.  Specifically
    it includes the modeling uncertainties associated with light-quark 
    fragmentation and out-of-cone corrections. For D\O\ \RunII\ measurements,
    it is included in the dJES category.
  \item[dJES:] That part of the JES uncertainty which originates from
   limitations in the data samples used for calibrations and which is
   correlated between measurements within the same data-taking
   period, such as \RunI\ or \RunII, but not between
   experiments.  For CDF this corresponds to uncertainties associated
   with the $\eta$-dependent JES corrections which are estimated
   using di-jet data events. For D\O\ this includes uncertainties in 
   the calorimeter response for light jets, uncertainties from 
   $p_{T}$- and $\eta$-dependent JES corrections and 
   from the sample dependence of using $\gamma$+jets data samples 
   to derive the JES.
  \item[rJES:] The remaining part of the JES uncertainty which is 
    correlated between all measurements of the same experiment 
    independently from the data-taking period, but which is uncorrelated between
    experiments.  For CDF, this is dominated by uncertainties in the
    calorimeter response to light-quark jets, and also includes small 
    uncertainties associated with the multiple interaction and underlying 
    event corrections. For D\O\ \RunII\ measurements, it is included in 
    the dJES category.
  \item[LepPt:] The systematic uncertainty arising from uncertainties
    in the scale of lepton transverse momentum measurements.  This is an
    important uncertainty for CDF's track-based measurement.  It was not
    considered as a source of systematic uncertainty in the \RunI\
    measurements. 
  \item[Signal:] The systematic uncertainty arising from uncertainties
    in the \ttbar\ modeling which is correlated across all
    measurements. This includes uncertainties from variations in the ISR,
    FSR, and PDF descriptions used to generate the \ttbar\ Monte Carlo samples
    that calibrate each method. For D\O\ it also includes the uncertainty 
    from higher order corrections evaluated from a comparison of \ttbar\ samples generated by {\sc MC@NLO} ~\cite{MCNLO} and 
    {\sc ALPGEN}~\cite{ALPGEN}, both interfaced to {\sc HERWIG}~\cite{HERWIG5,HERWIG6} for the simulation of parton showers and hadronization.
%
  \item[Background:]  Uncertainty in modeling the background sources. They are correlated between
    all measurements in the same channel, and include uncertainties on the background composition and shape.  In
    particular uncertainties associated with the modeling of the QCD
    multijet background using data in the all-j and l+j channels, uncertainties associated with the
    modeling of the Drell-Yan background in the di-l channel, and uncertainties associated 
    with variations of the factorization scale used to model $W$+jets 
    background are included.
  \item[Fit:] The systematic uncertainty arising from any source specific
    to a particular fit method, including the finite Monte Carlo statistics 
    available to calibrate each method. For D\O\ this uncertainty also includes 
    the uncertainties from modeling of the QCD multijet background determined 
    from data and dominated by limited statistics. 
  \item[Monte Carlo (MC):] The systematic uncertainty associated with variations
    of the physics model used to calibrate the fit methods and which is correlated
    across all measurements.  It includes variations observed when 
    substituting {\sc PYTHIA}~[37--39]
    (\RunI\ and \RunII) 
    or {\sc ISAJET}~\cite{ISAJET} (\RunI) for {\sc HERWIG}~\cite{HERWIG5,HERWIG6} when 
    modeling the \ttbar\ signal.  
  \item[Uranium Noise and Multiple Interactions (UN/MI):] This is specific to D\O\ and includes the uncertainty
    arising from uranium noise in the D\O\ calorimeter and from the
    multiple interaction corrections to the JES.  For D\O\ \RunI\ these
    uncertainties were sizable, while for \RunII, owing to the shorter
    calorimeter electronics integration time and {\em in situ} JES calibration, these uncertainties
    are negligible.
  \item[Color Reconnection (CR):] The systematic uncertainty arising from a variation of the 
  phenomenological description of color reconnection between final state 
  particles \cite{CR,Skands:2009zm}. This is obtained taking the difference between {\sc PYTHIA}\,6.4 tune ``Apro" and {\sc PYTHIA}\,6.4 tune ``ACRpro" that only differ only in the color reconnection model. Monte Carlo generators which explicitly include different CR models for hadron collisions have recently become available. This was not possible in Run I;  these 
measurements therefore do not include this source of systematic uncertainty.

  \item[Multiple Hadron Interactions (MHI):] The systematic uncertainty arising from a mismodeling of 
  the distribution of the number of collisions per Tevatron bunch crossing owing to the 
  steady increase in the collider instantaneous luminosity during data-taking. 
  This uncertainty has been separated from other sources to account for the fact that 
  it is uncorrelated with D\O\ measurements.

\end{description}
These categories represent the current preliminary understanding of the
various sources of uncertainty and their correlations.  We expect these to 
evolve as we continue to probe each method's sensitivity to the various 
systematic sources with ever improving precision.  Variations in the assignment
of uncertainties to the uncertainty categories, in the back-propagation of the bJES
uncertainties to \RunI\ measurements, in the approximations made to
symmetrize the uncertainties used in the combination, and in the assumed 
magnitude of the correlations, have all a negligible impact ($\ll 0.1\,\GeVc2$) in the 
combined \MT\ and total uncertainty.

\section{Correlations}
\label{sec:corltns}

The following correlations are used for the combination:
\begin{itemize}
  \item The uncertainties in the Statistical, Fit, and iJES
    categories are taken to be uncorrelated among the measurements.
  \item The uncertainties in the aJES, dJES, LepPt and MHI categories are taken
    to be 100\% correlated among all \RunI\ and all \RunII\ measurements 
    within the same experiment, but uncorrelated between \RunI\ and \RunII\
    and uncorrelated between the experiments.
  \item The uncertainties in the rJES and UN/MI categories are taken
    to be 100\% correlated among all measurements within the same experiment 
    but uncorrelated between the experiments.
  \item The uncertainties in the Background category are taken to be
    100\% correlated among all measurements in the same channel.
  \item The uncertainties in the bJES, cJES, Signal, CR, and MC
    categories are taken to be 100\% correlated among all measurements.
\end{itemize}
Using the inputs from Table~\ref{tab:inputs} and the correlations specified
here, the resulting matrix of total correlation coefficients is given in
Table~\ref{tab:coeff}.

\begin{table}[t]
\caption[Global correlations between input measurements]{The matrix of correlation coefficients used to determine the
  Tevatron average top-quark mass.}
\begin{center}
\renewcommand{\arraystretch}{1.30}
\begin{tabular}{l|ccccc|cc|cccc}
\hline \hline
     & \multicolumn{5}{c}{{\RunI} published} 
         & \multicolumn{2}{|c|}{{\RunII} published} 
         & \multicolumn{4}{c}{{\RunII} preliminary}   \\
      & \multicolumn{3}{c}{ CDF } 
         & \multicolumn{2}{c}{ D\O }
         & \multicolumn{2}{|c|}{ CDF } 
         & \multicolumn{2}{c}{ D\O } 
         & \multicolumn{2}{c}{ CDF } 
         \\
         
 &  l+j  &     di-l   &    all-j    &    l+j    &   di-l  &     all-j    &   trk    &    l+j    &     di-l    & l+j   & di-l  \\

\hline 
CDF-l+j     &   1.00   &		 -   &       -     &        -     &         -     &          -   &           -  &              -  &          -       &  -    &  - \\
CDF-I di-l     &   0.29   &   1.00   &      -      &      -      &         -      &       -      &        -     &           -     &       -         &  -    &  - \\
CDF-I all-j    &   0.32    &  0.19   &  1.00   &     -       &         -      &        -     &       -       &          -     &       -     &       -    &  - \\
D\O-I l+j      &   0.26   &   0.15   &   0.14  &    1.00  &        -      &      -       &         -      &        -      &       -      &     -    &  - \\
D\O-I di-l     &   0.11   &   0.08   &   0.07  &    0.16  &    1.00   &       -      &       -        &         -     &      -       &    -    &  - \\
CDF-II all-j    &   0.15   &   0.10   &   0.12  &    0.10  &    0.05   &   1.00   &       -         &      -       &   -        &     -    &  - \\
CDF-II trk    &    0.16   &   0.08   &   0.07 &     0.12  &    0.05   &   0.06  &    1.00     &       -      &     -      &    -    &  - \\
D\O-II l+j        &    0.10   &   0.08   &   0.06  &    0.07  &    0.04   &   0.10  &    0.11     & 1.00      &   -         &    -    &  - \\
D\O-II di-l         &   0.07   &   0.06   &   0.05  &    0.04  &    0.03   &   0.07   &   0.07     & 0.52    &  1.00     &    -    &  - \\
CDF-II l+j	    &   0.36   &   0.19   &   0.23    &   0.20	 &     0.07 &    0.19   &     0.19   &   0.22  &    0.15  &  1.00   &    -	\\
CDF-II  di-l       &   0.48  &   0.28   &   0.35    &   0.23	 &     0.11 &    0.21    &  0.12     &   0.11   &   0.08  &   0.43  &   1.00 \\

\hline
\hline
\end{tabular}
\end{center}
\label{tab:coeff}
\end{table}

The measurements are combined using a program implementing two independent methods: 
a numerical $\chi^2$ minimization and the analytic best linear unbiased estimator (BLUE) method~\cite{Lyons:1988, Valassi:2003}. 
The two methods are mathematically equivalent, and are also equivalent to the method used
in an older combination~\cite{TM-2084}. It has been checked that they give identical results for
the combination. The BLUE method yields the decomposition of the uncertainty on the Tevatron $\MT$ average in 
terms of the uncertainty categories specified for the input measurements~\cite{Valassi:2003}.

\section{Results}
\label{sec:results}

The combined value for the top-quark mass is:
%
$\MT=\gevcc{\measStatSyst{173.32}{0.56}{0.89}}$. 
Adding the statistical and systematic uncertainties
in quadrature yields a total uncertainty of $\gevcc{1.06}$, corresponding to a
relative precision of 0.61\% on the top-quark mass.
Rounding off to two significant digits in the uncertainty, the combination provides $\MT = \gevcc{173.3 \pm 1.1}$.
It has a $\chi^2$ of 6.1 for 10 degrees of freedom, which corresponds to
a probability of 81\%, indicating good agreement among all the input
measurements.  The breakdown of the uncertainties is 
shown in Table\,\ref{tab:BLUEuncert}. The total JES uncertainty is $\pm0.61\,$GeV/$c^2$ with $\pm0.46\,$GeV/$c^2$ coming 
from its statistical component and $\pm0.40$\,GeV/$c^2$ from the nonstatistical component.
The total statistical uncertainty is $\pm0.56\,$GeV/$c^2$.

\begin{figure}[p]
\begin{center}
\includegraphics[width=0.8\textwidth]{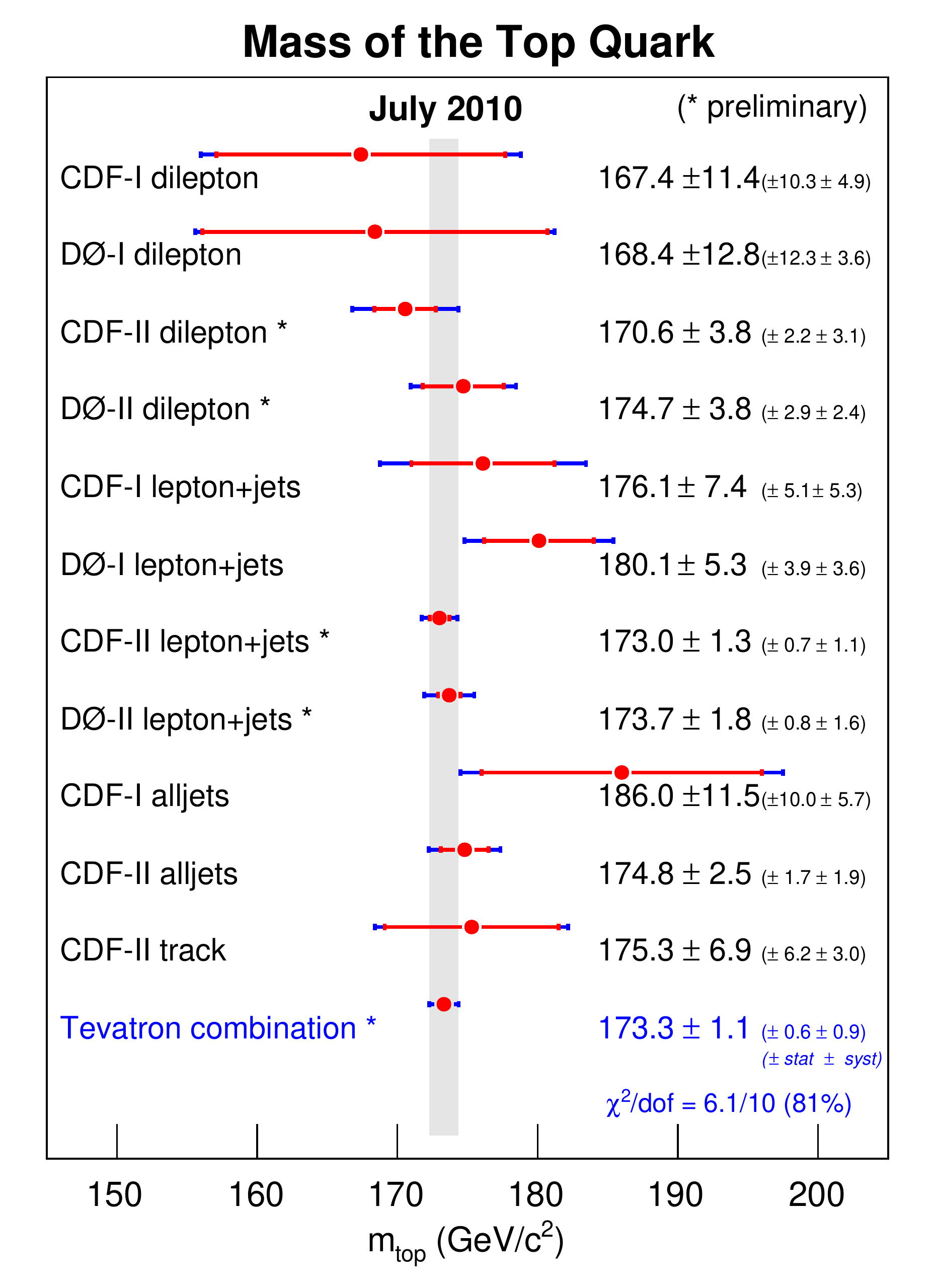}
\end{center}
\caption[Summary plot for the Tevatron average top-quark mass]
  {Summary of the input measurements and resulting Tevatron average
   mass of the top-quark.}
\label{fig:summary} 
\end{figure}

The pull and weight for each of the inputs are listed in Table~\ref{tab:stat}.
The input measurements and the resulting Tevatron average mass of the top 
quark are summarized in Fig.~\ref{fig:summary}.
\vspace*{0.10in}

\begin{table}[tbh]
\caption{\label{tab:BLUEuncert} 
Summary of the Tevatron combined average $\MT$ . The uncertainty categories are 
described in the text. The total systematic uncertainty and the total uncertainty are obtained 
by adding the relevant contributions in quadrature.}
\begin{center}
\begin{tabular}{lc} \hline \hline
  & Tevatron combined values (GeV/$c^2$) \\ \hline
$\MT$            & 173.32 \\ \hline
 iJES                     &    0.46 \\
 aJES                    &  0.21  \\
  bJES                   &  0.20 \\  
  cJES                   &  0.13 \\ 
  dJES                  &   0.19 \\ 
  rJES                   &   0.15 \\
  LepPt                 &  0.10 \\
  Signal                &   0.19  \\
  Background     &   0.23 \\
  Fit                       &   0.11 \\
  MC                     &   0.40 \\
  UN/MI                &    0.02 \\
  CR                     &   0.39 \\
  MHI                    &   0.08 \\ \hline
  Systematics      &   0.89 \\ 
  Statistics            &    0.56 \\ \hline
  Total                  &   1.06 \\
   \hline \hline
\end{tabular}
\end{center}
\end{table}

The weights of some of the measurements are negative. 
In general, this situation can occur if the correlation between two measurements
is larger than the ratio of their total uncertainties. This is indeed the case
here.  In these instances the less precise measurement 
will usually acquire a negative weight.  While a weight of zero means that a
particular input is effectively ignored in the combination, a negative weight 
means that it affects the resulting $\MT$ central value and helps reduce the total
uncertainty. 

\begin{table}[t]
\caption[Pull and weight of each measurement]{The pull and weight for each of the
  inputs used to determine the Tevatron average mass of the top quark.  See 
  Reference~\cite{Lyons:1988} for a discussion of negative weights.}
\begin{center}
\renewcommand{\arraystretch}{1.30}
{\small
\begin{tabular}{l|ccccc|cc|cccc}
\hline 
\hline
       & \multicolumn{5}{c}{{\RunI} published} 
       & \multicolumn{2}{|c|}{{\RunII} published} 
       & \multicolumn{4}{c}{{\RunII} preliminary}  \\
       & \multicolumn{3}{c}{ CDF } 
       & \multicolumn{2}{c}{ D\O\ }
       & \multicolumn{2}{|c|}{ CDF } 
       & \multicolumn{2}{c}{ D\O\ } 
       & \multicolumn{2}{c}{ CDF } \\
       & l+j     & di-l    & all-j   & l+j     & di-l    & all-j     & trk     & l+j   & di-l    & l+j   & di-l\\
\hline

Pull   & $+0.38$  & $-0.52$  & $+1.11$  & $+1.30$  & $-0.39$  & $+0.62$  & $+0.29$  & $+0.29$  & $+0.37$   & $-0.50$  & $-0.76$  \\
Weight [\%]  & $- 2.5$ & $- 0.5$ & $- 0.7$ & $+ 1.3$ & $+ 0.2$ & $+10.5$ & $- 0.5$ & $+ 26.2$ & $- 2.1$        
       & $+70.0$ & $-1.8$  \\
\hline \hline
\end{tabular}
}
\end{center}
\label{tab:stat} 
\end{table} 


Although no input has an anomalously large pull and the
$\chi^2$ from the combination of all measurements indicates
that there is good agreement among them, it is still interesting to also fit for the top-quark mass
in the all-j, l+j, and di-l channels separately.  We use the same methodology,
inputs, uncertainty categories, and correlations as described above, but fit 
the three physical observables, \MTjj, \MTlj, and \MTll separately.
The results of these combinations are shown in Table~\ref{tab:three_observables}.

Using the results of Table~\ref{tab:three_observables} we calculate the chi-squared consistency
between any two channels, including all correlations, as 
$\chi^{2}$(di-l/l+j)=0.75, $\chi^{2}$(l+j/all-j)=0.68, and 
$\chi^{2}$(all-j/di-l)=1.58.  These correspond to chi-squared probabilities of 38\%, 68\%, and 21\%, respectively, and indicate 
that the determinations of \MT\ from the three different final states are consistent with each another.

%
%

\begin{table}[t]
\caption[Mtop in each channel]{Summary of the combination of the 11
measurements by CDF and D\O\ in terms of three physical quantities,
the mass of the top quark in the all-j $\MTjj$, l+j $\MTlj$, and di-l $\MTll$ decay channels. }
\begin{center}
\renewcommand{\arraystretch}{1.30}
\begin{tabular}{ccrrr}
\hline\hline
Parameter & Value (\GeVc2) & \multicolumn{3}{c}{Correlations} \\
               &                                 & $\MTjj$ &    $\MTlj$  &  $\MTll$ \\ \hline
$\MTjj$ & $175.2\pm 2.6$    & 1.00       &                   &       \\
$\MTlj$ & $173.0\pm 1.1$    & 0.20       &    1.00       &      \\
$\MTll$ & $171.1\pm 2.5$    & 0.19        &    0.48      & 1.00 \\
\hline\hline
\end{tabular}
\end{center}
\label{tab:three_observables}
\end{table}

\section{Summary}
\label{sec:summary}

A preliminary combination of measurements of the mass of the top quark
from the Tevatron experiments CDF and D\O\ is presented.  The
combination includes five published \RunI\ measurements, two published \RunII\ measurements, and 
four preliminary \RunII\ measurements.  Taking into
account the statistical and systematic uncertainties and their
correlations, the preliminary result for the Tevatron average is: 
  $\MT=\gevcc{\measStatSyst{173.32}{0.56}{0.89}}$,
where the total uncertainty is obtained  assuming Gaussian systematic uncertainties.
Adding in quadrature the statistical and systematic uncertainties
yields a total uncertainty of $\gevcc{1.06}$, corresponding to a
  relative precision of 0.61\% on the top-quark mass. Rounding off the uncertainty to two significant digits, the combination provides $\MT = \gevcc{173.3 \pm 1.1}$.

The central value is 0.20\,GeV/$c^2$ higher than our March 2009 average of $\MT=173.12\pm1.26$\,GeV/$c^2$, while the relative precision
has improved by 15\% with respect to the previous Tevatron average.
\vspace*{0.10in}

The mass of the top quark is now known with a relative precision of
0.61\%, limited by the systematic uncertainties, which are dominated by
the jet energy scale uncertainty.  This source of systematic uncertianty is expected to
improve as larger datasets are collected since analysis
techniques constrain the jet energy scale using kinematical information from $W\ra
qq^{\prime}$ decays. It can be expected that with the full
\RunII\ dataset the top-quark mass will be known to an accuracy better than
the one presented in this paper.  To reach this level of precision further work will 
be focuses on a better understanding of $b$-jet modeling, and in the uncertainties in the signal and 
background simulations.

\section{Acknowledgments}
\label{sec:ack}

We thank the Fermilab staff and the technical staffs of the
participating institutions for their vital contributions. 
This work was supported by  
DOE and NSF (USA),
CONICET and UBACyT (Argentina), 
CNPq, FAPERJ, FAPESP and FUNDUNESP (Brazil),
CRC Program, CFI, NSERC and WestGrid Project (Canada),
CAS and CNSF (China),
Colciencias (Colombia),
MSMT and GACR (Czech Republic),
Academy of Finland (Finland),
CEA and CNRS/IN2P3 (France),
BMBF and DFG (Germany),
Ministry of Education, Culture, Sports, Science and Technology (Japan), 
World Class University Program, National Research Foundation (Korea),
KRF and KOSEF (Korea),
DAE and DST (India),
SFI (Ireland),
INFN (Italy),
CONACyT (Mexico),
NSC(Republic of China),
FASI, Rosatom and RFBR (Russia),
Slovak R\&D Agency (Slovakia), 
Ministerio de Ciencia e Innovaci\'{o}n, and Programa Consolider-Ingenio 2010 (Spain),
The Swedish Research Council (Sweden),
Swiss National Science Foundation (Switzerland), 
FOM (The Netherlands),
STFC and the Royal Society (UK),
and the A.P. Sloan Foundation (USA).

\clearpage

\providecommand{\href}[2]{#2}\begingroup\raggedright\endgroup



\end{document}